\documentclass[12pt]{article}
\newcommand\ignore[1]{}
\usepackage{float}
\usepackage[left]{lineno}
\usepackage{amsmath}
\usepackage{amssymb}
\usepackage{cancel}
\usepackage{graphicx}
\usepackage{braket}
\usepackage{authblk}
\usepackage{caption,subcaption}
\usepackage{comment}
\usepackage{enumitem}
\usepackage{color}

\usepackage[margin = .75 in]{geometry}
\usepackage[pdftex, pdfstartview={FitH}, pdfnewwindow=true, colorlinks=false, pdfpagemode=UseNone]{hyperref}

\graphicspath{{fig/}}

\newcommand\be{\begin{equation}}
\newcommand\ee{\end{equation}}
\newcommand\bea{\begin{eqnarray}}
\newcommand\eea{\end{eqnarray}}

\usepackage[final]{showlabels}

\title{ 
		\vskip 30pt
		Topological charge unfreezing with AMReX}
	
	\author{Dean Howarth$^a$ and Adam J Peterson$^{a,b}$
		\vspace{0 cm}\\
		\normalsize {\it $^{a}$ Center for Computational Science and Engineering}, \\
        {\it Lawrence Berkeley National Laboratory}, Berkeley, CA 94720 USA
        \vspace{0.3 cm}\\
		\normalsize {\it $^{b}$ Strategic Deterrence Directorate}, \\
        {\it Lawrence Livermore National Laboratory}, Livermore, CA 94550 USA
		\vspace{.3 cm}\\
		{ \footnotesize peterson129@llnl.gov}  
	}

\begin{document}
\newpage

\date{}

\maketitle

\begin{abstract}
A new approach to the problem of topological freezing in gauge theories is introduced in which a physical volume preserving coarsening of the lattice induces sufficient energy variation in the Hamiltonian to overcome large topological barriers. Though the process is not reversible, the physical volume preserving aspect minimises the time spent rethermalisating the lattice after coarsening periods, which we then treat as a new ensemble disjoint from previous runs. We have tested this technique on the pure gauge 2D Schwinger model and find that topological sampling rates are improved. We also demonstrate that autocorrelation times for extensive observables are restored to their pre-coarsening values after coarsening periods over a acceptably short simulation time.
\end{abstract}

\setlength{\parskip}{0in}
\thispagestyle{empty}
\setcounter{page}{-1}
\pagebreak
\tableofcontents
\thispagestyle{empty}
\setcounter{page}{0}
\pagebreak
\setcounter{page}{1}
\setlength{\parskip}{.2in}

\section{Introduction}

Lattice field theories use Markov Chain Monte Carlo (MCMC) generation of field configurations and importance sampling to measure expectation values of observables. This program has shown excellent agreement with experimental observations in the case of lattice quantum chromodynamics (LQCD) \cite{Workman:2022ynf}. As simulations employ finer lattice spacing, there is an increase topological barrier sizes that prohibit simulations from properly sampling all topological sectors, a phenomenon known as `topological freezing'. It has been suggested by Luscher \cite{Luscher:2017cjh,Luscher:2011kk} to use a large physical volume `master field' in which extremely large lattices with open temporal boundary conditions permit sufficient topological transitions, but such a technique is naturally constrained by the sheer size of the lattice required. A variety of other update algorithms exist which address this problem, such as collective variable metadynamic \cite{Laio:2015era}, instanton updates \cite{Dilger:1994ma}, and (pertinent to this discussion) multiscale thermalisation \cite{Endres:2015yca, Detmold:2016rnh, Detmold:2018zgk}.

An interesting proposal by Foreman {\it et al} \cite{Foreman:2021rhs} uses machine learning techniques to increase the $\beta$ value of the simulation during an HMC Leapfrog step in a 2D Schwinger $U(1)$ gauge theory in an exactly reversible fashion, though this technique does not yet permit the inclusion of dynamical fermions.

In this paper we propose a solution to topological freezing that utilizes selective, volume preserving coarsening of moderately sized lattices. This targeted coarsening allows the lattice simulation to escape from a topological minima and, upon refinement to the original scale, thermalise in a new sector. Targeting a restricted volume of the lattice for coarsening allows for quick rethermalization times and restoration of autocorrelation of observables, as opposed to a full volume re-gridding which can have more profound effects on autocorrelation.

To this end we make use of AMReX \cite{Zhang2019}, a software framework for massively parallel adaptive mesh refinement (AMR) applications, to provide selective coarsening of the lattice. In particular, we apply an adaptive mesh coarsening (AMC) procedure to a two dimensional Schwinger model and illustrate the effects on both the topological charge measurements, as well as a selected set of observables.  AMReX also provides the user with many options for interpolation and averaging operations between the coarse and fine levels, allowing for flexibility in the parameters.  For this initial demonstration of we will consider only a few options for simplicity.

The presentation is organized as follows: we begin with a review of the lattice 2D Schwinger model in section 2, as well as a discussion of the definitions of observables we use in the procedure to follow. In section 3 we introduce the basics of the AMReX framework and our modifications to the standard HMC procedure making use of AMC techniques.  In section 4 we illustrate our results.  Section 5 addresses avenues of future research, with concluding remarks and discussion in section 6.

\section{Lattice Schwinger Model}

To illustrate our approach to AMC unfreezing we make use of the euclidean 2D Schwinger model.  For ease of demonstration we will not include fermions in this initial exploration. This will also simplify the process of coarsening, which we will consider in the next section.

For this case the lattice Schwinger model, the pure gauge action is defined as:
\begin{equation}
    S = \beta \sum_{{\bf n} \in \Lambda} {\rm Re} \left(  1 -  \Pi({\bf n})\right),
\label{LatticeAction}
\end{equation}
where the plaquette $\Pi$ is defined as:
\begin{equation}
    \Pi({\bf n}) = U_1({\bf n})U_2({\bf n} + a{\bf e}_1)U^{\dag}_1({\bf n} + a{\bf e}_2)U^{\dag}_2({\bf n}),
\label{Plaguette}
\end{equation}
where ${\bf e}_{1,2}$ are unit vectors.
The continuum limit of (\ref{LatticeAction}) is:
\begin{equation}
    S = \frac{\beta a^2}{2}\int d^2x F_{\mu\nu}^2, \;\; F_{\mu\nu} = \partial_\mu A_\nu -\partial_\nu A_\mu,  \;\; \beta = \frac{1}{(ea)^2}
\label{ContinuumAction}
\end{equation}
for the dimensionless gauge field $A_\mu$.  Here $a$ is the lattice spacing, which is included to make the coupling $\beta$ dimensionless. The charge coupling $e$ mass dimension is 1 in the Schwinger Model. We will also make use of the phase angle of the gauge link:
\begin{equation}
    \theta_\mu \equiv a A_\mu.
\end{equation}
These choices will ease the discussion of coarsening in the following section.

The Schwinger model admits global instanton sectors given by the topological charge:
\begin{equation}
    n = \frac{1}{2\pi}\int d^2x F_{\mu\nu}{\Tilde{F}}_{\mu\nu},
\label{ContinuumInstanton}
\end{equation}
which at finite lattice spacing gives:
\begin{equation}
    n = \frac{1}{2\pi}\sum_{N \in \Lambda} \arg(\Pi).
\end{equation}

\section{Methods of adaptive mesh coarsening}

Here we describe our approach and packages used to achieve adaptive mesh coarsening on the 2-D lattice. We make use of the well-known AMReX framework with modifications for the coarsened-oriented methods required for lattice field theory problems. 

\subsection{AMReX and modifications}

\begin{figure}
	\centering
        \begin{subfigure}{0.44\textwidth}
        \centering
        \includegraphics[width=1.0\linewidth]{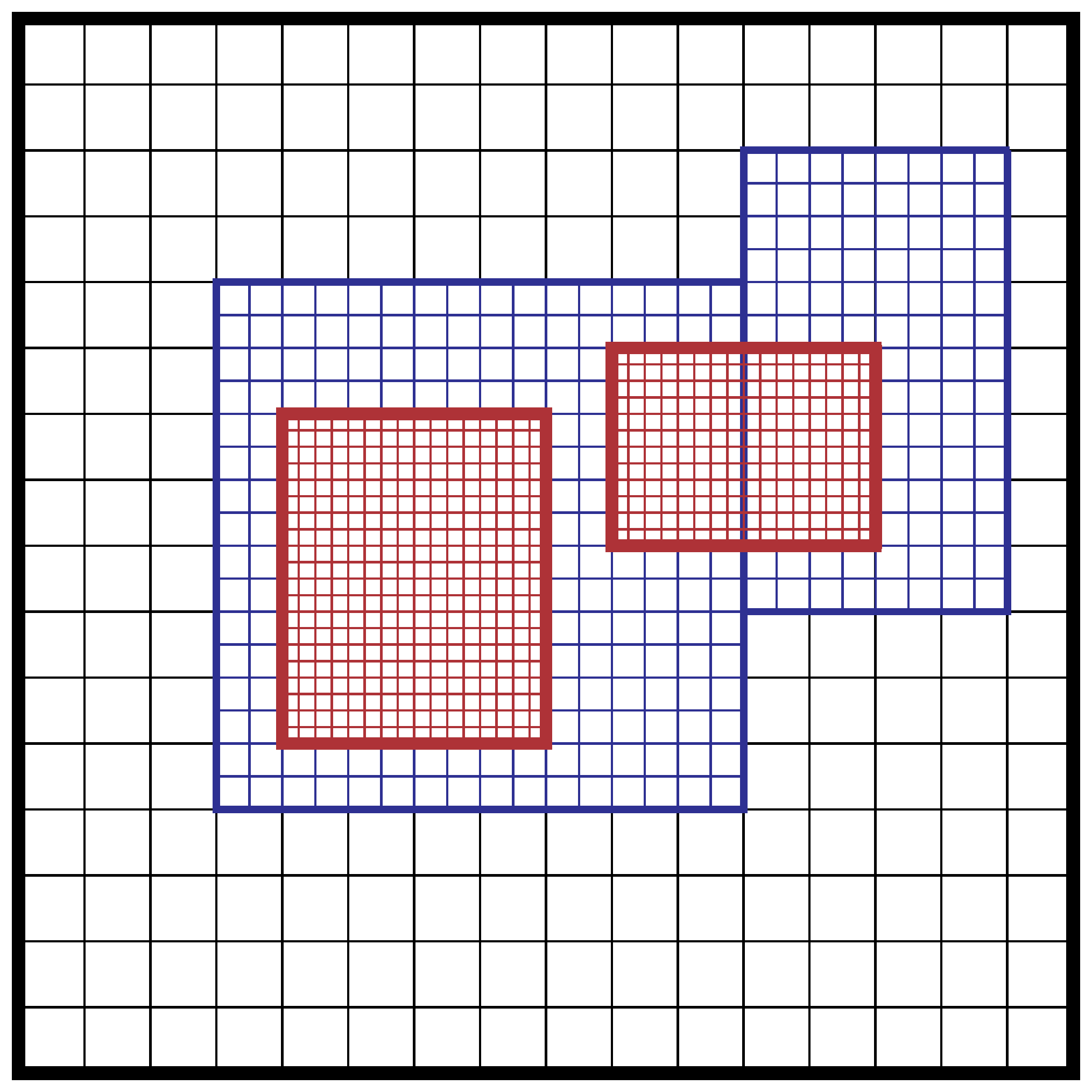}
	\caption{Standard AMR}
	\label{AMRGridsIllustration_a}
        \end{subfigure}%
        \hspace{1cm}
	\begin{subfigure}{0.45\textwidth}\label{fig1b}
        \centering
        \includegraphics[width=1.0\linewidth]{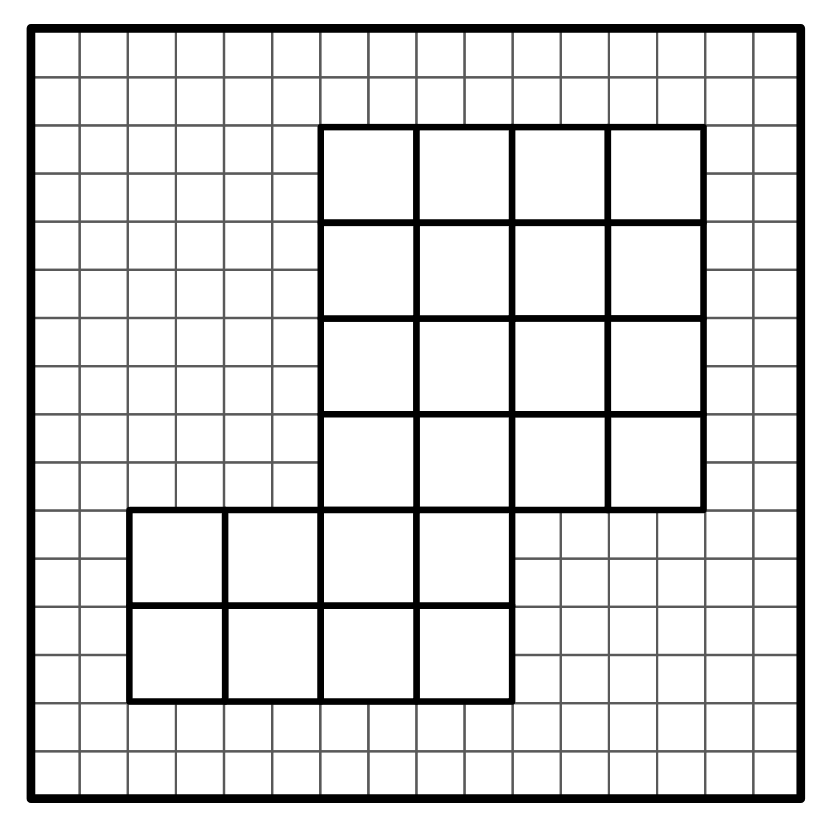}
	\caption{AMC}
	\label{AMRGridsIllustration_b}
        \end{subfigure}
        \caption{Shown on the left is the standard block structured AMR procedure, where higherlevels are refined and nested within lower coarser levels.  We use a modified procedure (right) where the coarser levels are considered higher levels nested within a refined domain.}
        \label{AMRGridsIllustration}
\end{figure}

We have chosen to use the AMReX software framework for developing our adaptive mesh modified HMC procedure for the lattice Schwinger model.  AMReX is a software framework for providing massively parallel, block-structured adaptive mesh refinement (AMR) applications.  AMReX uses a nested hierarchy of logically rectangular grids, an example of which is illustrated in Figure \ref{AMRGridsIllustration_a}. Refined grid domains are always bounded by their parent grid domain, though they may intersect multiple blocks within that parent grid.  Each grid block is formulated with appropriate ghost cells which are either filled with boundary conditions (if the block intersects the domain boundary) or by interpolation from the next coarser grid.

AMReX supports both cell centered data as well as nodal data in any or all of the directions.  However, since AMReX is optimized for cell centered data we select this option, though in future research, nodal data for the gauge links is likely the more efficient choice.  For this initial probe of this problem, the specific data type will have little impact on the conclusions.  We will elaborate on this point in the discussion sections below. In our case we have modified our AMReX based code to act as an inverted version of the AMR procedure, shown in Figure \ref{AMRGridsIllustration_b}, where we consider the lowest level to be the finest level and consider coarsened regions as nested patches within the full domain.  

\subsection{AMC version of lattice Schwinger} 

Consider the dimensionless field $\theta_\mu$ and dimension-1 field $\pi_\mu$ as our dynamical variables, where $\pi_\mu$ is the conjugate momentum to $\theta_\mu$. On each level we must evolve $\pi_\mu$ using the correct length scale $a$.  We may ignore the length scale dependence of $\beta$ in this case, since the Schwinger model receives no anomalous corrections in the absence of fermions. To be more concrete, if we set the length scale of a level to be $a_F = a$, then the next coarsest level length scale may be defined as:
\begin{equation}
a_C = 2 a_F.
\end{equation}

Written in terms of $\theta_\mu$ and $\pi_\mu$ our HMC hamiltonian is written as:
\begin{equation}
    H_{\rm HMC} = \int d^2x \left\{\ \frac{\pi_\mu^2}{2}  + \frac{\beta}{2}\left(\partial_{[\mu} \theta_{\nu]}\right)^2\right\} \rightarrow \sum_{n \in \Lambda}a^2 \frac{\pi_\mu^2}{2}+S,
\label{HMC_Hamiltonian}
\end{equation}
where S is the lattice action (\ref{LatticeAction}). Note that the conjugate momentum $\pi_\mu$ has mass dimension 1 in this setup.

From this Hamiltonian we can determine the equations of motion for $\pi_\mu$:

\begin{equation}
    \dot{\pi}_\mu = - \frac{\delta S}{\delta \theta_\mu} =  2\frac{ \beta}{a^2}\hat{\partial}_\nu\left( \hat{\partial}_{\nu}\theta_\mu - \hat{\partial}_{\mu}\theta_\nu\right),
    \label{PiEOM}
\end{equation}
where we define a discrete derivative operator $\hat{\partial}_\mu$
\begin{equation}
    \hat{\partial}_\mu f \equiv f(n + \hat{\mu}) - f(n) = a\partial_\mu f.
\end{equation}
Note that $\dot{\pi}_\mu$ refers to the Monte-Carlo time derivative of $\pi_\mu$.
This form illustrates the length scale dependence of the equations of motion for $\pi_\mu$.  In particular, when evolving (\ref{PiEOM}) we must take care that the factor $a^{-2}$ is appropriately chosen for the grid level.

\subsection{AMC algorithm for the 2-D lattice Schwinger model}

As in standard approaches to MCMC based LQCD we generate a sequence of field configurations using the HMC method of perturbing the conjugate momenta and evolving the field configuration in a molecular dynamics evolution for a chosen amount of substeps to produce a new field configuration.  This new configuration is either accepted or rejected via the Metropolis procedure with weighted likelihood of acceptance $r = \min(1,\exp(-\Delta E/T))$, for an energy change $\Delta E$ between the initial and final field configurations.

The new ingredient in our procedure is to terminate the Markov Chain by coarsening grids between two (or more) levels. Once a sufficient number of configurations from one topological sector on the fine lattice are collected, we halt data collection, arbitrarily choose lattice links to be coarsened, average the fine gauge link values to the new coarse link values and then resume evolution using standard HMC with coarsened links. We suspend the accept reject step for the duration of the existence of coarsened links, and accept all new configurations. Following enough HMC steps to introduce sufficient topological noise to induce a transition to another sector, the grid is refined back to the original levels, gauge link values for these fine links are interpolated from the coarse link values and the HMC algorithm resumes for a period of rethermalisation. Upon sufficient thermalisation, the metropolis step is reintroduced and data collection is resumed.

It is clear that reversibility is lost during transitions between coarse and fine grids due to the gauge link value interpolation. However, this does not affect the fidelity of the collected data. We consider the simulation after the coarsening procedure to be a new ensemble, starting from partially thermalised state, and thermalise the new ensemble.

Ergodicity is also lost due to this coarsening procedure, and the topological charge distribution of conglomerated ensembles for these Markov chains will not be identical to that of a single Markov chain simulation. However, one may use a reweighing of the collected ensembles to maintain the proper distribution of topological charge sectors in the full sample. The topological charge is normally distributed with mean 0, hence only the width of the distribution (topological susceptibility) governs the form of the distribution. For our study, the proper distribution is estimated by extrapolating the topological susceptibility from purely fine simulations at strong coupling (low beta) where topological freezing is not as pronounced.

Our methods for averaging and interpolating between levels are as follows.  We define the following function on the lattice:

\begin{equation}
    F({\bf n}) \equiv \sum_{\mu} ({\bf n}_\mu \; {\rm mod } \; 2).
\end{equation}

I.e., for every lattice point whose {\it x,y} coordinates are both even, $F({\bf n})$ will evaluate to 0. Using this definition, our method for defining links on a coarser level from the fine level data will be defined as:
\begin{equation}
    U^c_\mu\left(\frac{{\bf n}}{2}\right) = U^f_\mu({\bf n})U^f_\mu({\bf n} + {\bf e}_\mu),
\end{equation}
 and similarly for the momenta,
\begin{equation}
    \pi^c_\mu\left( \frac{{\bf n}}{2} \right) = \pi^f_\mu ( {\bf n} + n_1 {\bf e}_1 + n_2 {\bf e}_2),
\end{equation}
where $n_{1,2}$ are randomly chosen as $-1$, $0$, or $1$, and $F({\bf n}) = 0$,

For interpolating from the coarse level to the fine level we set:
\begin{equation}
    U^f_\mu ({\bf n})= U^c_\mu\left(\frac{{\bf n}}{2}\right), \;\; U^f_\mu({\bf n} + {\bf e}_\mu) = 1,
\end{equation}
and
\begin{equation}
    \pi^f_\mu({\bf n}) = \pi^c_\mu\left(\frac{{\bf n}}{2} + n_1 {\bf e}_1 + n_2 {\bf e}_2\right), \; \pi^f_\mu({\bf n}+{\bf e}_\mu) = \pi^c_\mu\left(\frac{{\bf n}}{2} + n_1 {\bf e}_1 + n_2 {\bf e}_2\right)   
\end{equation}
where $n_{1,2}$ are randomly chosen integers between $-2$ and $2$, and $F({\bf n}) = 0$.

For nodal data these methods preserve the topological charge between levels, although we consider cell centered data here.  We reiterate that the specific method for interpolating/averaging is irrelevant when thermalization takes place between measurements.

\section{Results}

Here the finest level will consist of a $n_x = n_y = 16$ cell-centered lattice of lattice spacing $\Delta x = a n_x$, and $\Delta y = a n_y$. We choose cell-centered data, as AMReX is optimized in this case.  We will discuss the benefits of node-centering for gauge links in subsequent sections.

We perform each simulation for $500,000$ pure gauge Leapfrog HMC steps of trajectory length $\tau=1.0$ with $12$ integration substeps for the molecular dynamics evolution.  We set the coupling $\beta = 7.0$, chosen beyond the point of observed topological freezing for this lattice size. This allows us to compare results between standard HMC procedures, and our AMC modified approaches. We consider four cases of the HMC method, including the standard grid with no AMC steps (though including periodic rethermalisation to compare with the AMC cases which require these steps), the case with full domain coarsening, and the case of grid fracturing, with $2\times 2$ and $4 \times 4$ cell tagging for coarsening. The procedure is run with periodic grid fracturing taking place every 200 trajectories. The grids will be evolved in the fractured state for 2 trajectories and then interpolated to the finest level, with a further 18 steps for rethermalising (where the MCMC accept/reject step is suspended). The remaining 180 steps proceed as in the standard HMC method.

\subsection{Behavior of Topological Charge}
\begin{figure}
	\centering
    \begin{subfigure}{0.5\textwidth}
    \centering
	\includegraphics[width=1.0\linewidth]{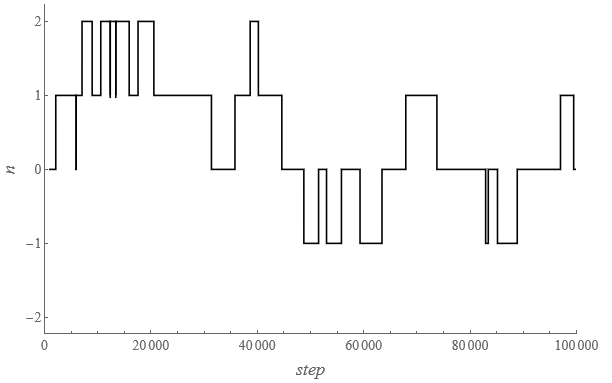}
    \end{subfigure}%
    \begin{subfigure}{0.5\textwidth}
    \centering
    \includegraphics[width=1.0\linewidth]{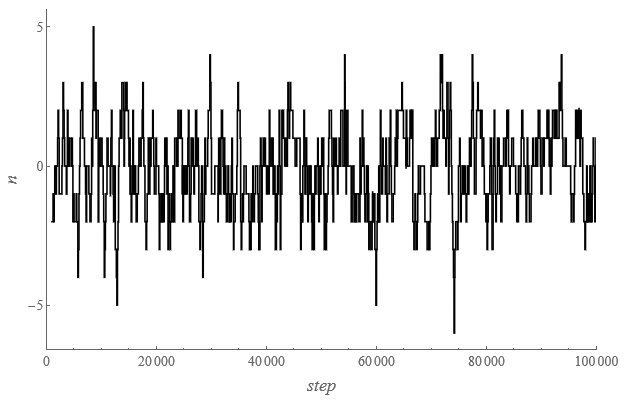}
    \end{subfigure}
    \begin{subfigure}{0.5\textwidth}
    \centering
	\includegraphics[width=0.95\linewidth]{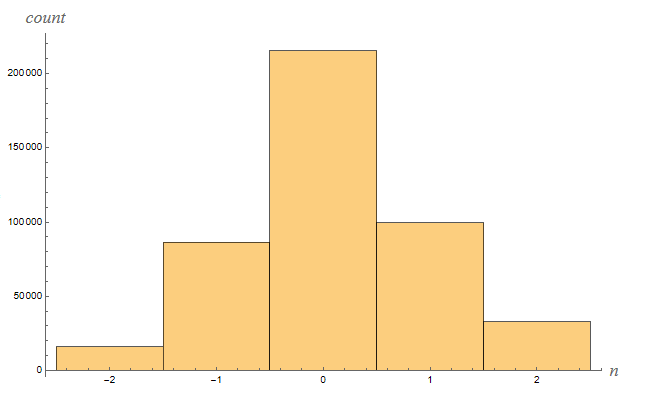}
    \end{subfigure}%
    \begin{subfigure}{0.5\textwidth}
    \centering
    \includegraphics[width=.95\linewidth]{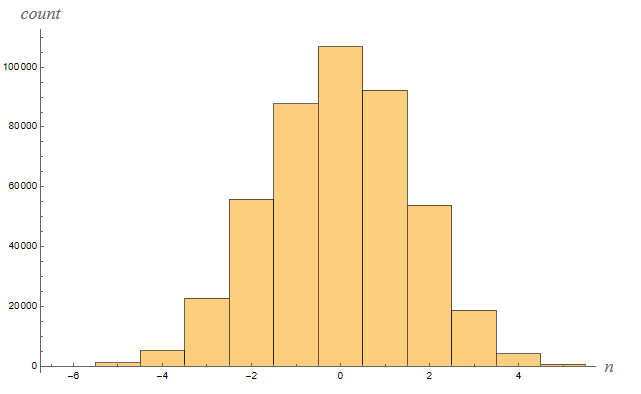}
    \end{subfigure}
	\caption{A comparison of the topological charge evolution for standard HMC (top,left), and with periodic AMC coarsening of a $2\times2$ set of cells (top, right). Additionally, the associated histograms of binned counts for topological charge cases are shown for the standard case (bottom, left), and the AMC case (bottom, right).}
\label{TopChargeVsStep}
\end{figure}

Figure \ref{TopChargeVsStep} shows the raw behavior of the topological charge as the simulation is evolved for both the standard case, and the $2\times2$ AMC procedure discussed above. Similar behavior is observed for the full domain AMC and $4\times4$ AMC cases, where the procedure of coarsening significantly increases variance in the topological charge.  It is also apparent that the topological charge distribution produced using AMC has a larger variance (susceptibility) than is expected when using standard methods without any coarsening.

\begin{figure}
    \centering
	\includegraphics[width=0.8\linewidth]{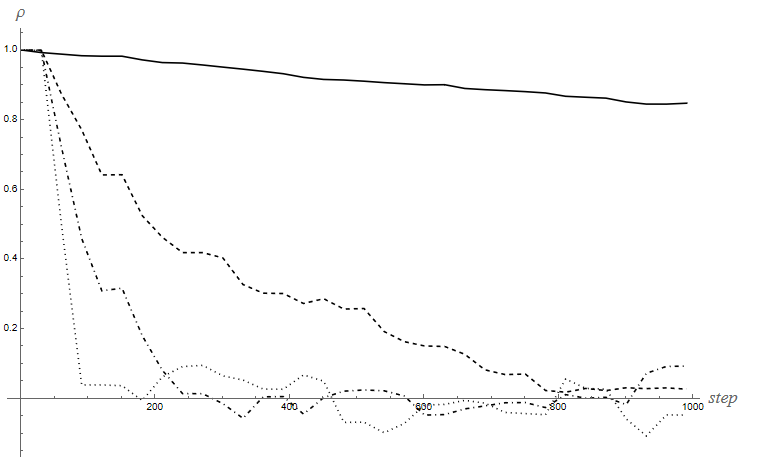}
	\caption{Topological charge autocorrelation versus the simulation step count.  The case of no AMC coarsening is shown in black, $2\times2$ AMC coarsening as dashed, $4\times4$ AMC coarsening as dash-dotted, and the full grid AMC coarsening as dotted.}
	\label{TopChargeAutocorrelation}
\end{figure}

Figure \ref{TopChargeAutocorrelation} shows the measured autocorrelation $\rho$ for the topological charges for all cases of AMC for $\beta = 7.0$.  We see that all of the AMC methods tested result in shorter autocorrelation times for the topological charge.

To correct for this artificially inflated distribution we sample the variance of the topological charge distributions versus coupling $\beta$.  We sample using lower values of $\beta$ where the variance in topological charges does not suffer from freezing.  We then extrapolate to the case of $\beta \ge 6.0$ in the region of critical slowdown, and apply a correction to the measured distribution when AMC is included.

\begin{figure}
	\centering
    \begin{subfigure}{0.48\textwidth}
    \centering
	\includegraphics[width=1.0\linewidth]{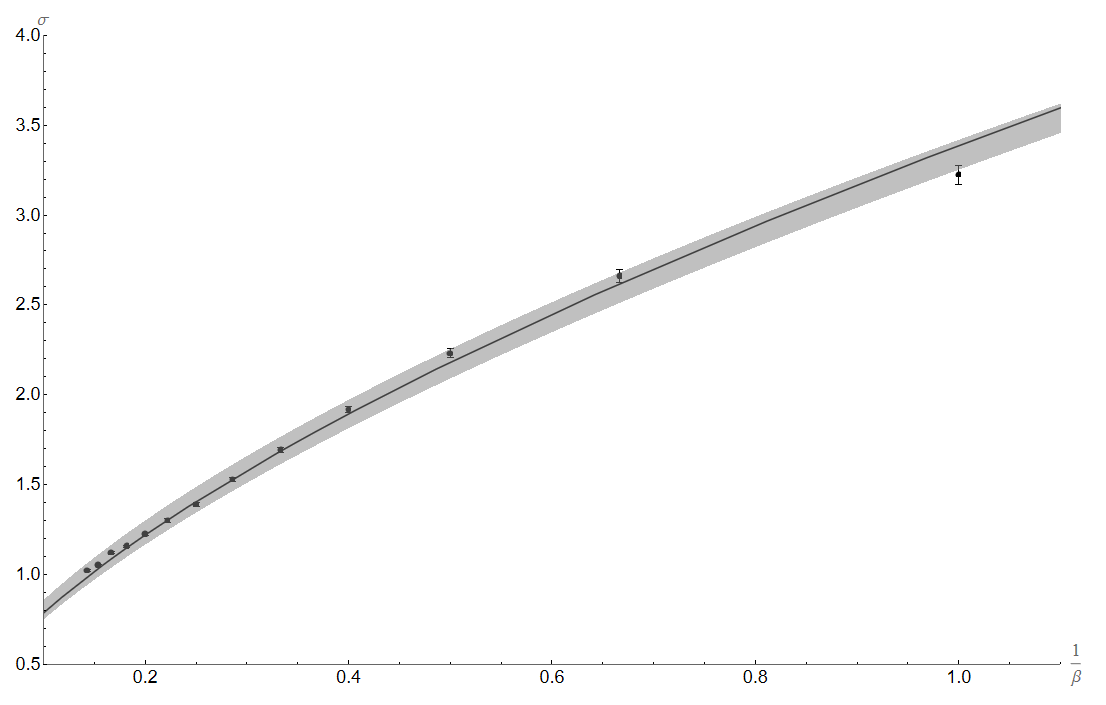}
    \end{subfigure}%
    \begin{subfigure}{0.5\textwidth}
    \centering
    \includegraphics[width=1.0\linewidth]{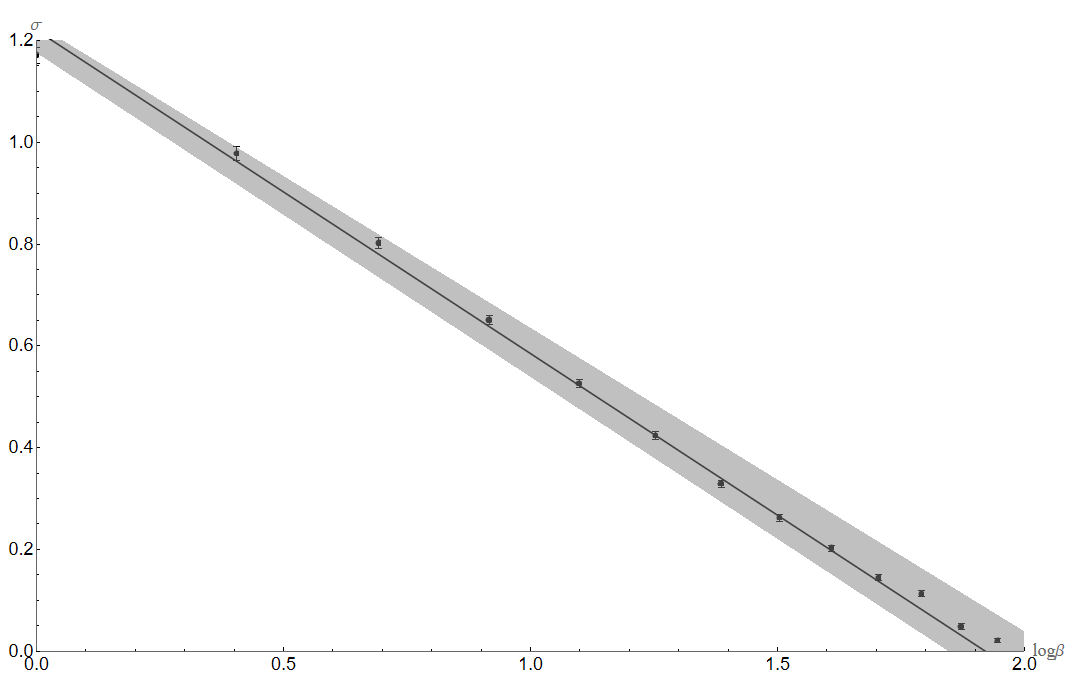}
    \end{subfigure}
    \caption{The measured value of the topological susceptibility $\sigma$ versus the coupling strength $\beta$. This allows us to extrapolate $\sigma$ into regions of large beta so that we may produce a sound topological charge distribution for ensembles produced with the coarsening procedure. Error bars are determined from the 4th moment of topological charge distributions.  Error bars are doubled for visibility.}
    \label{SigmaVsBeta}
\end{figure}

The resulting behavior of the variance $\sigma$ is shown in figure \ref{SigmaVsBeta} where the fit is determined for the values of $\beta \le 4.0$.  We include measurements of $\beta \ge 4.0$ to illustrate the extrapolation for large values of the coupling.

\subsection{Action and Wilson loop autocorrelation}

\begin{figure}
    \centering
	\includegraphics[width=0.83\linewidth]{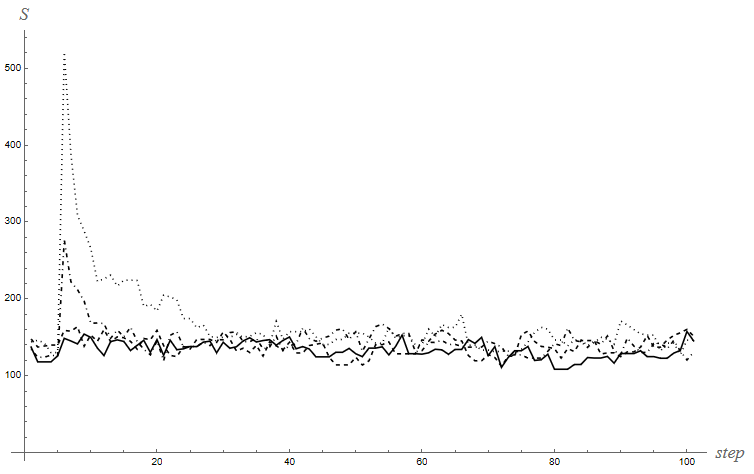}	
	\caption{Gauge action after 2 trajectories of coarse evolution and 18 trajectories of rethermalisation. Black = No AMR. Dashed = 2x2 AMR.  Dot-dashed = 4x4 AMR.  Dotted = Full Domain AMR.}
	\label{ActionComparison}
\end{figure}

\begin{figure}
	\centering
    \begin{subfigure}{0.5\textwidth}
    \centering
	\includegraphics[width=1.0\linewidth]{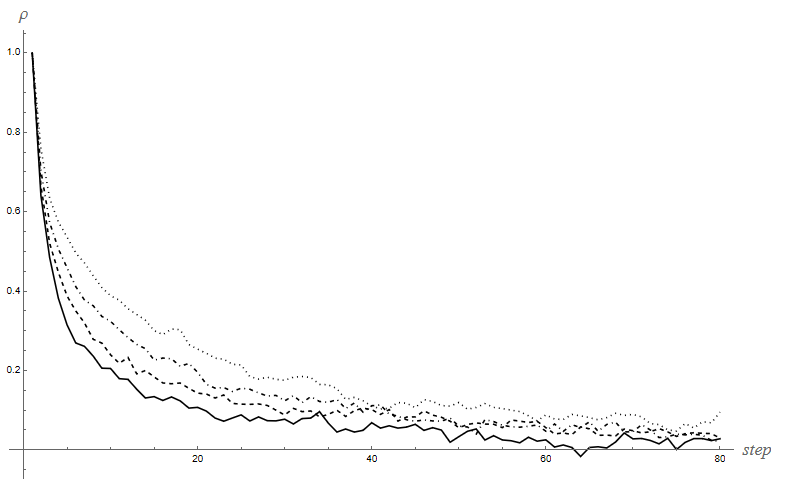}
    \end{subfigure}%
    \begin{subfigure}{0.5\textwidth}
    \centering
    \includegraphics[width=1.0\linewidth]{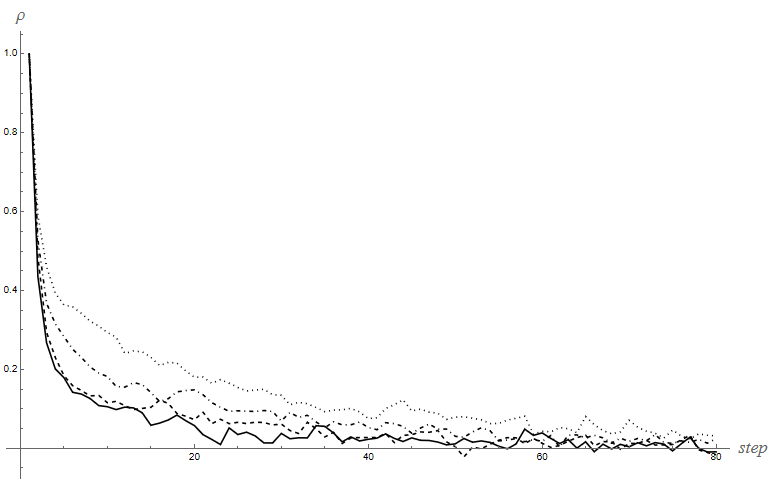}
    \end{subfigure}
    \caption{Autocorrelations of the $1\times1$ (left) and $4\times4$ Iright) Wilson loops after 2 trajectories of coarse evolution and 18 trajectories of rethermalisation. Black = No AMR. Dashed = 2x2 AMR.  Dot-dashed = 4x4 AMR.  Dotted = Full Domain AMR.}
    \label{WilsonLoopComparison}
\end{figure}

The utility of the AMC method is further demonstrated in figures \ref{ActionComparison} and \ref{WilsonLoopComparison}.  In figure \ref{ActionComparison} we show a measurement of the gauge action following the AMC procedure, and compare this with the gauge action measurements from a standard simulation with no coarsening.  Unsurprisingly, we observe improved rethermalization of the action for the cases of $4\times4$ and $2\times2$ target coarsening.  This is verified with the measured autocorrelations of the $1\times1$ Wilson loops shown for all cases in figure \ref{WilsonLoopComparison}.  For smaller regions of targeted coarsening, autocorrelations steadly approach those of the standard case.  This result is further supported for larger Wilson loops as shown for the $4\times4$ Wilson loop autocorrelation in figure \ref{WilsonLoopComparison}.

\section{Avenues of future research and improvements}

\subsection{Improving AMC methods}

In order to improve the method of adaptive coarsening it is suggested that future research consider octree approaches to AMR as opposed to the block structured approach in AMReX.  In this project we modified AMReX to allow for an inverted data level system to consider the finest data on the lowest level, with coarser data as descendants. In order to achieve this significant modifications were made to the cell tagging procedures to consider finer cells as tagged for coarsening (as opposed to the standard method of tagging coarse cells for refinement).  As octree formulations are not bounded by the block structuring, performance would be improved in this case. 

It is also worth noting that we have only considered only a very simplified version of coarsening, with coarsening taking place periodically at fixed locations on the lattice. One may also explore the use of machine learning to identify particular regions of the lattice targeted for coarsening. One could train a model to identify which (minimal) number of links should be coarsened to induce a topological charge transition. The model may reveal patterns dependent on topological charge density, such as links which connect sites where the charge denisty undergoes a sign change, or local regions of the lattice where the charge density sign is homogeneuous.


\subsection{Cell centered data versus nodal data}
Throughout this simulation and analysis, we have used cell centered data for all lattice variables.  As AMReX is optimized for cell centered data this was the logical choice for this project.  Additionally, rethermalisation allows for flexibility when considering different data types and interpolation/averaging procedures.  

However, we must point out that lattice simulations are most meaningful when gauge links are considered node-to-node on the cell edges.  This makes immediate sense from the definition of gauge links.  However, the process of lifting link data from a fine level to the coarser level is only well defined when the links are defined on cell edges.  Strictly speaking, in the case of cell centering, a mismatch between point locations occurs when matching data between levels.  For our analysis this effect can alter the measured topological charges on a fractured grid.  To reduce rethermalisation times, future projects should be built on nodal grids.

\subsection{Towards lattice QCD and fermions}

For this project we made use of the 2D lattice Schwinger model without dynamical fermions in place of a full 4 dimensional 3 flavour lattice QCD model, with many simplifying assumptions and procedures to make the demonstration of and AMC approach to HMC evolution more approachable. To elevate this approach to LQCD, the use of 4 spatial dimensional AMR package will have to be developed in place of AMReX's limited 3-dimension support. Alternatively, existing LQCD application codes would have to generate sub lattices which hold coarse level link data, and augment their code to use the sub lattice data during `coarsening' periods. 

Finally we have opted to exclude dynamical fermions for this initial investigation to simplify the coarsening procedure. In future research we will include dynamical fermions, which will be straightforward, and thus analysis will be required to consider the anomalous corrections to the coupling $\beta$. When performing coarsening and refinement to the lattice, care will have to be taken to ensure the physical lattice spacing on each level is tuned to keep physical observables as consistent as possible. For example, when two fine links at $\beta_{fine}$ are coarsened into a single link, the $\beta_{coarse}$ value at that link should be chosen such that the lattice spacing is twice as large as the fine link spacing. 

\section{Summary and concluding remarks}

We have presented a new method using adaptive mesh coarsening procedures to increase the variability of long range observables susceptible to critical slow down for small lattice spacing (large $\beta$). We have specifically demonstrated the utility of this approach for the case of topological charge freezing in the pure $U(1)$ gauge 2D Schwinger model. We find that for the observables we considered, adaptive coarsening has the effect of reducing autocorrelation times for the topological charge, while maintaining expected autocorrelation times for the gauge action and Wilson loops after rethermalisation.

In distinction to other proposed update techniques and algorithms, this technique avoids large `master field' type simulations  and allows for the use of moderate lattice sizes. Further, we mitigate the loss of reversibility in our algorithm by treating each instance of coarsening as a break in the MCMC chain, thus each subsequent simulation as a new ensemble. Relinquishing reversibility constraints makes the inclusion of dynamical fermions straightforward. We address ergodicity issues by extrapolating measured topological susceptibility to regions of large $\beta$ so that a sound statistical distribution can be constructed with adequate weighting of configurations by topological charge. We also avoid excessive rethermalisation by targeting only local regions of the lattice rather than its entirety, and we propose techniques to reduce the burden of rethermalisation even further.

\vspace{1.0cm}

\section*{Acknowledgements}
This research was funded by the Exascale Computing Project (17-SC-20-SC).  The authors are very thankful to Andre Walker-Loud, and Ann Almgren for useful discussions and advice for this project.

\bibliography{references}
\end{document}